\documentclass[lettersize,journal,onecolumn]{IEEEtran}
\usepackage{amsmath,amsfonts}
\usepackage{algorithmic}
\usepackage{algorithm}
\usepackage{array}
\usepackage{textcomp}
\usepackage{stfloats}
\usepackage{url}
\usepackage{verbatim}
\usepackage{graphicx}
\usepackage{cite}
\usepackage{xcolor}
\usepackage{tikz}
\usepackage{subcaption}
\usepackage{todonotes}
\usepackage{balance}

\usetikzlibrary{shapes.geometric, arrows.meta, positioning}
\newcommand{\airisp}{AIRIS$^\mathrm{2}$}
\newcommand{\airis}{AIRIS$^\mathrm{2}$~}
\newcommand{\mbf}{\mathbf}
\newcommand{\event}{e_g^{\Delta_t,\alpha}(t)}
\newcommand{\eventh}{\hat{e}_g^{\Delta_t,\alpha}(t)}

\newcommand{\jcm}[1]{{\color{black}{#1}}}
\newcommand{\jcmm}[1]{{\color{black}{#1}}}

\newcommand{\sizing}[2]{#1}

\begin{document}

\title{\airisp : a Smart Gateway Diversity Algorithm \\ for Very High-Throughput Satellite Systems}

\author{Justin Cano,
		Jonathan Israel 
		and Laurent Feral
\thanks{The authors are with DEMR, ONERA, Universit\'e de Toulouse, 31000 Toulouse, France.  $\{$\texttt{jcano,jisrael,lferal}$\}$\texttt{@onera.fr}.}
\thanks{This study has been founded within the ONERA/DSG PRORAD program.}
\thanks{Preprint version}}

\markboth{Preprint Version}%
{Cano \MakeLowercase{\textit{et al.}}: \airis : a Smart Gateway Diversity Algorithm \\ for Very High-Throughput Satellite Systems}


\maketitle

\begin{abstract}
\jcmm{Satellite communication systems are shifting to higher frequency bands (Ka, Q/V, W) to support more data-intensive services and alleviate spectral congestion.} 
\jcmm{However, the use of Extremely High Frequencies, typically above 20 GHz, causes significant tropospheric impairments, such as rain attenuation, which can causes system outages.}
To mitigate these effects, Smart Gateway Diversity (SGD) has emerged as a promising method for maximizing feeder link availability through an adaptive site diversity scheme. However, implementing such technique requires a decision-making policy to dynamically select the optimal set of gateways and prevent outages.
\jcm{This paper introduces \airisp, a deep learning algorithm that anticipates short-term rain events from \jcmm{rain}
	attenuation measurement to enable efficient gateway switching.} 
	\jcmm{The approach is validated from}
	  five years of measured time series
	  \jcmm{collected at Ka and Q/V bands at various sites and climatic conditions.} 
\end{abstract}


\section{Introduction}

\IEEEPARstart{V}{ery} High Throughput Satellite (VHTS) systems are increasingly demanding bandwidth to support high 
services such as video streaming or low-ping banking applications. 
To meet this demand, manufacturers tend to increase carrier frequencies to EHF bands (such as the Q/V band, above $40~\mathrm{GHz}$) \jcmm{to achieve high data rates} and avoid spectral congestion \cite{cianca_ehf_2011}.

However, at such frequencies, tropospheric attenuation  can reach tens of decibels, and therefore lead to system outages \cite{boulanger_propagation_2019,garcia-del-pino_joint_2010,suquet_twelve_2024,ventouras_assessment_2021}. 
These impairments are
mainly due to rainfall that causes absorption \jcmm{and scattering of the incident radioelectric wave}. The magnitude of these phenomena 
\jcmm{increases with the frequency and the raindrop size}
	\cite{alozie_review_2022}. 

Despite these physical constraints, the availability of the VHTS systems must be maximized to provide high-reliability services (\textit{e.g.}, available more than $99.9\%$ of the time) \cite{maral_satellite_2020}. Therefore, Fade Mitigation Techniques (FMT) have to be implemented to counteract these impairments. In particular, the \jcmm{\textit{feeder links} of VHTS systems}, \textit{i.e.}, from the Gateway Earth Stations (GES) to the satellite, are the most critical to meet this availability requirement \cite[Chap. 5.3]{arbesser-rastburg_cost_2002}. Indeed, an outage on these links could potentially deny the service to thousands of users. Moreover, feeder links are generally established \jcmm{at a higher carrier frequency in order to meet} their capacity \jcmm{requirements} \cite{maral_satellite_2020} that makes them more vulnerable to rain fading \cite{arbesser-rastburg_cost_2002,kyrgiazos_terabitsecond_2014,alozie_review_2022,suquet_twelve_2024}.
%
	\jcm{Therefore, one intuitive FMT known as the UpLink Power Control (ULPC) consists in the dynamic adjustment of the GES transmission power, which aims to counteract tropospheric attenuation events up to a few decibels \cite{arbesser-rastburg_cost_2002}.}
In addition, digital processing techniques such as Adaptive Modulation and Coding (ACM) can \jcmm{mitigate slight atmospheric impairments by increasing the robustness of the Modulation and Coding (ModCod) chain}.
\jcmm{However, this strategy implies a reduction in the spectral efficiency, which is detrimental to the Quality of Service (QoS) \cite{ebert_method_2020,maral_satellite_2020}.} 
 
\begin{figure}[h]
	\centering
	\includegraphics[width=\sizing{0.4}{0.8}\linewidth]{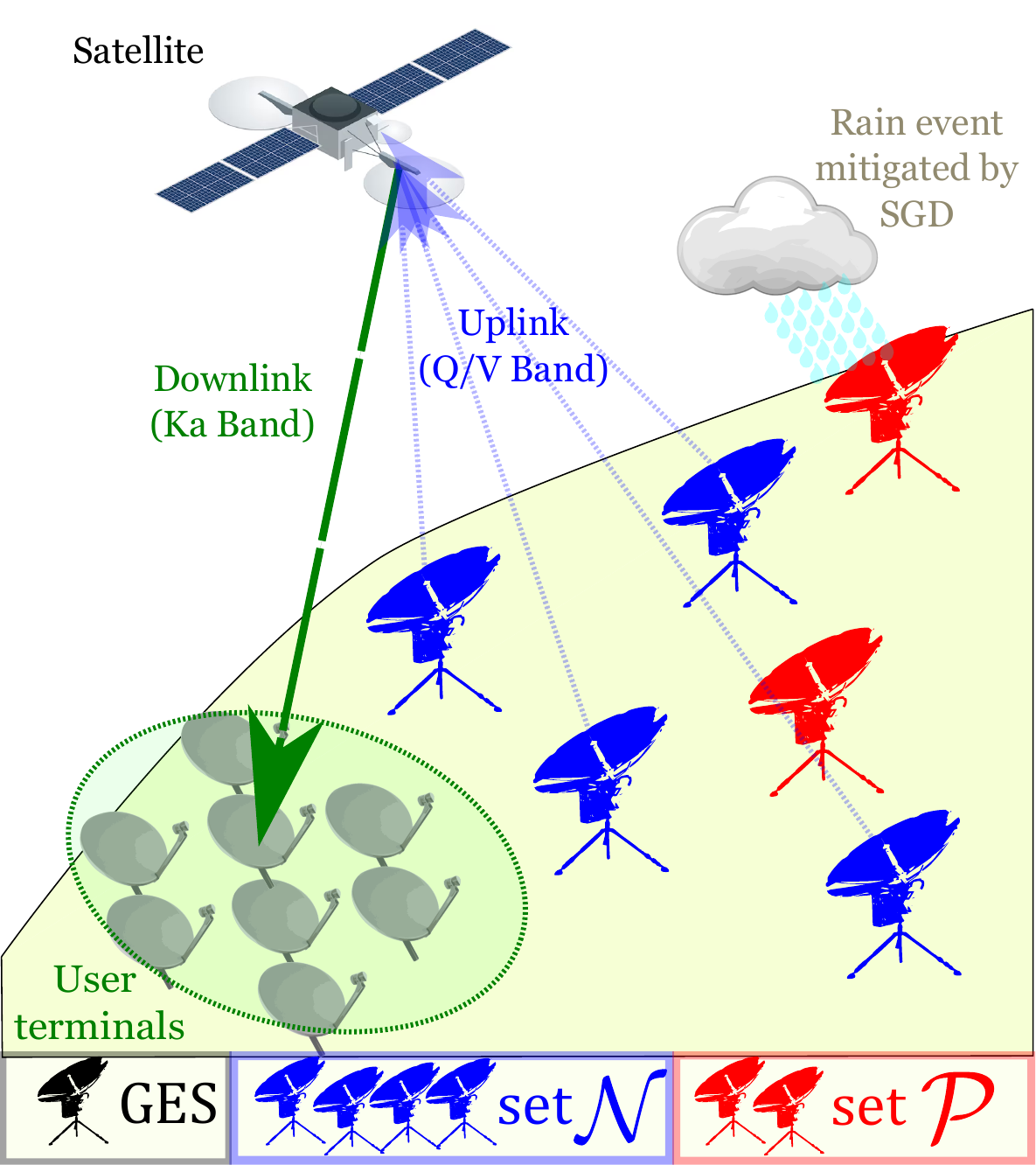}
	\caption{SGD concept for VTHS systems.}
	\label{fig:sgd}
\end{figure}

To counteract strong fade phenomena, Smart Gateway
Diversity (SGD) is a promising FMT that acts on the uplink of the VHTS system  \cite{jeannin_smart_2019}. This feeder link originates from the GES. The idea behind SGD is to have two sets of gateways: an active set $\mathcal{N}$ of $N$ GES, used when tropospheric propagation conditions are \jcmm{favorable,}
and a backup set $\mathcal{P}$ of $P$ GES, used to mitigate fade events. $N$ and $P$ are constant  but the GES composing the two sets varies over time \cite{jeannin_smart_2019,ventouras_assessment_2021}. An illustration of such concept is shown in Fig. \ref{fig:sgd}, where the sets $\mathcal{N}$ and $\mathcal{P}$ are in blue and red, respectively, and where a GES from the back-up set $\mathcal{P}$ is mitigating a rain event.

\paragraph*{Notation} Throughout this paper, we denote in bold vectors and matrices and the estimates with a hat. Time indexes are noted between parenthesis, agent index as subscripts, while parameterization is noted as superscript. The probability of a given statistical event $e$ is denoted as $\mathbb{P}[e]$. 
%

\section{Problem statement}
\label{sec:statement}

\jcm{Here, we are interested in the development of SGD policies that are the algorithms controlling the switches of GES, \textit{i.e.} replacing an impaired GES
$n\in \mathcal{N}$ by a certain $p\in \mathcal{P}$, which \jcmm{is} in favorable propagation conditions to maximize the system availability \cite{jeannin_smart_2019}.}
\subsection{Related Work}

Since the role of SGD policies is to prevent system outages, most of the related literature focuses on the prediction of these events. In particular, a probabilistic modeling of the outages is given in \cite{efrem_computation_2021}, relying on the Poisson binomial distribution and validated with simulations. In \cite{rossi_smart_2020}, Rossi \textit{et al.}
propose a methodology to optimize the configuration of a SGD based on a similar statistical modeling but including sparsity and network constraints in the programming.
Overall, these references assess an optimal SGD system design, though not a real-time GES switching policy when the network \jcmm{is already deployed}.

The optimization of SGD from a network system perspective is also investigated in \cite{aurizzi_sdn_2019}, where the allocation of Software Defined Networks (SDN) is discussed for the design of SGD. On the other hand, the work by Roumeliotis \textit{et al.} \cite{roumeliotis_optimized_2021} addresses this challenge using sophisticated constraint-based programming, known as the Hungarian algorithm, which tackles both network allocation and outage minimization problems. These studies have been validated through simulations based on strong assumptions about channel behavior that are difficult \jcm{to fulfill in practice}.

Our approach tends to be close \jcmm{to} propagation, directly working on excess attenuation measurements from Geostationary Earth Orbit (GEO) satellites carrying beacons (\textit{e.g.} Alphasat). \jcmm{We assume that the system is outaged if the attenuation exceeds a given level, in compliance with \cite{kyrgiazos_gateway_2014} and \cite{ventouras_assessment_2021}.}
These approaches consist in switching \jcmm{a GES $n$ from $\mathcal{N}$ to $\mathcal{P}$ 
whenever the instantaneous attenuation $a_n(t)$ exceeds a given threshold $\alpha$, \textit{i.e.}, $a_n(t)>\alpha$}.
However, in practice, the switching action requires backbone network operation which can last a few minutes, resulting in the appearance of a system delay $\Delta_t$ \cite{ventouras_assessment_2021}. Therefore, \jcmm{this} direct threshold approach \jcmm{relies on} a \textit{Persistence Hypothesis} (PH), assuming the constancy of the attenuation between the decision taken at time $t$ and the effective switching at time $t+\Delta_t$. 

However, neglecting such delay, which can be up to a couple of minutes \cite{ventouras_assessment_2021}, necessarily results in False Negative (FN) decisions (undetected outages) that harms the QoS. Finally, \jcmm{it should be noted} that the previous papers were only validated over simulations, \jcmm{because} of the difficulty in collecting a significant amount of synchronous data \jcmm{across} a complete VHTS system. \jcmm{In addition}, none of them includes short-term prediction of the channel state to compute switching decision.

\subsection{Contribution}

Here, we restrain to the case \jcmm{where we only have} the uplink excess attenuation time series measurements $\{a_g(k)\}_{k\in[0,t];\forall g \in \mathcal{N} \times \mathcal{P}},$ at disposal to take a switching decision at time $t$. This is motivated by the \jcmm{need to develop}
low-cost switching policies, without any costly extrinsic measurement\jcmm{s} such as meteorological radar. We also assume that there exists a tens-of-seconds delay $\Delta_t$, which makes the switching effective at $t+\Delta_t$. Another constraint of the problem, \jcmm{underlined} in \cite{ventouras_assessment_2021}, is to have a reasonable number of SGD switches. \jcmm{Indeed, a frequent variation of the network topology provoked by the SGD} could harm the performances of the network routing protocol. 

Overall, an operative SGD algorithm should anticipate the switching decision at time $t+\Delta_t$ while minimizing False Negative (FN) decisions to save the QoS from outages and maintaining a few False Positive (FP) to prevent erroneous switching. In this paper, we propose a SGD \jcm{policy} that tackles these issues, by predicting severe attenuation at time $t+\Delta_t$ leveraging neural networks. Our algorithm, called \textit{AI-Rain Impairment SGD Switcher} (\airisp) as a tribute to the upcoming European mega-constellation IRIS$^2$, is detailed in Section \ref{sec:algo}.

Since \airis  is AI-powered, it highly depends on the input data used to train it. Therefore, we detail the datasets used in \ref{ss:data}; the training procedure and computational time are specified in \ref{ss:train}. 
Finally, \airis is tested over several standard use-cases and compared to the baseline persistence algorithm. The tests include both \jcmm{geographic}, frequency and parametric variations for the models learned by \airis and are presented in Section \ref{sec:results}.

\section{$\mathrm{AIRIS^2}$ Algorithm}
\label{sec:algo}
The goal of \airis is to predict the boolean events $\event:= a_g(t+\Delta_t)>\alpha$ for a certain GES \jcmm{$g \in \mathcal{N}\cup \mathcal{P}$} based on a measured attenuation vector $\mathbf{a}_{g}(t)$, made with available data at time $t$, with the delay $\Delta_t$ and the threshold $\alpha$, supposed known. 
In this Section, we detail the implementation of such algorithm, the input data $\mathbf{a}_{g}(t)$ and the training process.

\subsection{An LSTM-based architecture}
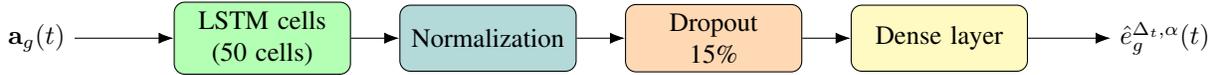
\begin{figure*}[t]
	\centering 
	\begin{tikzpicture}[node distance=3cm, >=Latex] 
		\tikzstyle{block} = [rectangle, draw, fill=#1!30, text width=6em, text centered, minimum height=2.5em, rounded corners] \tikzstyle{line} = [draw, -{Latex[length=2mm]}]
		\node (input) {\jcmm{$\mathbf{a}_{g}(t)$}}; 
		\node [block=green, right of=input] (lstm) {LSTM cells \\ (50 cells)}; 
		\node [block=teal, right of=lstm] (normalization) {Normalization}; 
		\node [block=orange, right of=normalization] (dropout) {Dropout \\ 15\% }; 
		\node [block=yellow, right of=dropout] (dense) {Dense layer}; 
		\node [right of=dense, node distance=3cm] (output) {$\eventh$}; 
		\path [line] (input) -- (lstm); 
		\path [line] (lstm) -- (normalization); 
		\path [line] (normalization) -- (dropout); 
		\path [line] (dropout) -- (dense); 
		\path [line] (dense) -- (output); 
	\end{tikzpicture}	
	\caption{\airis predictor architecture.}
	\vspace{-5mm}
	\label{fig:archi}
\end{figure*}

Our algorithm is based on an architecture that uses Long Short Term Memory (LSTM) as main component. These AI elementary bricks, developed in the late 90s \cite{hochreiter_long_1997}, are nowadays very common in the machine learning literature \cite{chollet_deep_2021}. In particular, they are used to predict meteorological events \jcm{in the} short-term, \textit{e.g.} rainfall in \cite{bhimavarapu_irf-lstm_2022} or wind in \cite{liu_short-term_2021}. 
Overall, the LSTM cells are acting like memory blocks that can highlight time dependencies within a given time series. However, rather than solving a regression problem (\textit{i.e.} predicting $\hat{a}_{g}(t+\Delta_t)$), we aim to predict the Boolean variable $\event$ based on $\mbf a_g (t)$. \jcmm{This consists in a classification problem that motivates the use of a dense layer activated by a sigmoid function, which computes the probability of the tropospheric event to come $\event$.} 

An additional layer of normalization that consists in scaling the data to an unit-centered Gaussian distribution is leveraged. This process aims to enhance the learning algorithm convergence, which relies on stochastic gradients \cite{chollet_deep_2021}. Finally, a dropout layer that randomly masks $15\%$ of the data for the gradient computation at each iteration, is added in order to minimize the learning bias, \textit{i.e.}, the over-fitting phenomenon \cite{srivastava2014dropout}. 
The architecture used is summarized in Fig. \ref{fig:archi} and is implemented using the Keras Python library \cite{noauthor_keras_nodate}.

\subsection{\jcmm{Experimental dataset}} 
\label{ss:data}

The \jcmm{input data} used are excess attenuation measured over \jcmm{the} years, sites and frequencies reported in Table \ref{tab:data}. \jcmm{The multi-years time series have been} processed using the methodology explained in \cite{boulanger_four_2015} and sampled at $10~\mathrm{Hz}$. Overall, four representative datasets are \jcm{available} at two frequencies (Q/V and Ka bands) in tropical (Kourou and Hassan) and temperate (Toulouse) \jcmm{areas}, which allows \airis to be validated in various climates, \jcm{mirroring the conditions of a realistic} SGD system.

\begin{table}[h]
	\caption{Input experimental excess attenuation data.}
	\label{tab:data}
	\centering
	\begin{tabular}{|c|c|c|c|c|}
		\hline
		\jcmm{GES} & Location & Frequency & Years & Satellite \\
		\hline
		$T_1$ & Toulouse, France & 40GHz & 2018-2022 & Alphasat \\
		\hline
		$T_2$ & Toulouse, France & 20GHz & 2018-2022 & Astra 3B \\
		\hline
		$K$ & Kourou, F. Guiana & 20GHz & 2016-2021 & Amazonas 3 \\
		\hline
		$H$ & Hassan, India & 20GHz & 2015-2019 & GSAT-14  \\
		\hline
	\end{tabular}
\end{table}

The data is sliced in input vectors, for each time $t$, $\mathbf{a}_g(t):=[\dots a_g(l)\dots] , l\in[t-H,t]$, where $H$ is the history of the vector and $g\in\{T_1,T_2,K,H\}$ being the considered \jcmm{gateway} in Table \ref{tab:data}. For each $t$, we compute the associated event $\event$. Finally, we gather the input vector $\mbf{X}_g=[ \dots ,  \mbf a_g(t) , \dots ]$ and the (output) ground truth  vector $\mbf y_g = [\dots, \event ,\dots]$ . The length $H$ must be chosen with respect to the horizon of prediction $\Delta_t$, here, we have chosen to tune the history length abiding by the rule $H:=2\Delta_t$. 

\subsection{Training Process}
\label{ss:train}

The collected data \jcmm{are} randomly separated into three databases: a training database ($70\%$ of the data) used for the learning process, a validation database ($15\%$) to remove bias during the learning and an independent test database ($15\%$) that is used to compute the performances of our predictor, which is a classical strategy in deep learning \cite{chollet_deep_2021}. Since the rain events are scarce, \textit{e.g.} $\mathbb{P}[a(t)>15\mathrm{dB}] \approx 3\%$ for $T_1$, we chose to balance the training and validation databases in order to enhance the convergence of the learning algorithm. Namely, we set $\mathbb{P}[\event=1]=\mathbb{P}[\event=0]$ for $t$ in training or validation databases. Indeed, the prediction problem solved by \airisp, see as a classification problem allows \jcm{for such balancing, which is more challenging} for time-series regression problems with rare events.       

\begin{figure}[h]
	\centering
	\includegraphics[width=\sizing{0.8}{1.0}\linewidth]{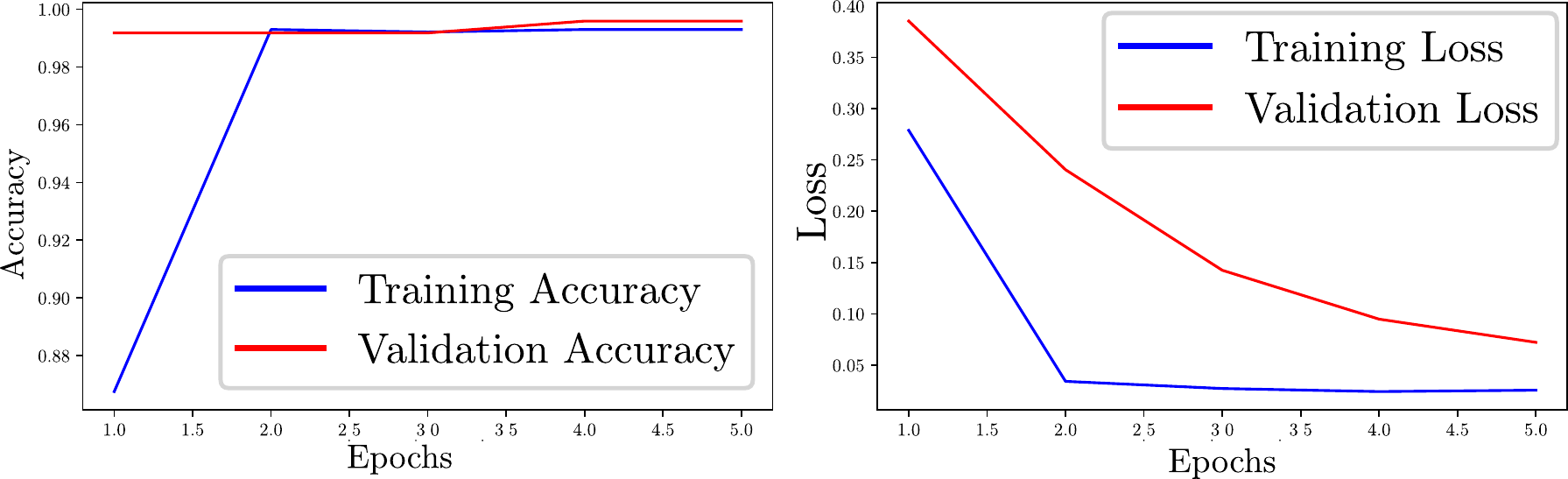}
	\caption{Learning curves (accuracy and cost function) for $\alpha=15\mathrm{dB}$ and $\Delta_t=60~s$ for $T_1$ dataset.}
	\label{fig:learning}
\end{figure}

In order to train \airis, we leverage the Adam stochastic gradient-based optimizer \cite{kingma_adam_2017}, using five epochs with 64-length batches. Since the problem assessed by \airis is a boolean classification, \textit{i.e.}, $\event \in \{0,1\}$, the optimization was performed on the binary cross-entropy cost function \cite{mannor_cross_2005}. Therefore, we use the accuracy metric which gives the empirical probability $\mathbb{P}(\event=\eventh; t\in \mathcal{D})$ over a given database $\mathcal{D}$. An example of learning process is detailed in Figure \ref{fig:learning} that shows an optimization process (minimization of the cost function) over learning and validation databases, alongside the accuracy metric for a given parameterization $(\alpha,\Delta_t)$. This figures highlights the efficiency of learning, since the accuracy is high (around $99\%$) after only two epochs of learning. Overall, \airis training time took less than 10 minutes for each pairs $(\alpha,\Delta_t)$ and databases presented in Section \ref{ss:data}, and was performed using a 112 processor Linux server. Nonetheless, this figure depends on the amount of data and the attenuation threshold $\alpha$ (since the number of events decreases if $\alpha$ increases) provided at the input of the learning process but seems reasonable \jcm{given} the reduced number of parameters ($1051$) to train. 

\newcommand{\eventph}{\hat{e}_{g,\mathsf{PH}}^{\Delta_t,\alpha}(t)}
\section{Experimental Results}
\label{sec:results}

In this section, we present an experimental validation of \airis over the datasets presented in Table \ref{tab:data}. 
We are interested in comparing $\eventh$ obtained by \airis to the prediction obtained by the PH predictor.
Namely, the PH gives immediately  
\[\eventph:=a_n(t)>\alpha\]
which assumes that the channel state is frozen during $\Delta_t$.
As aforementioned in Section \ref{sec:statement}, our work is focusing on the decrease of the FN rate, which could decrease the VHTS availability. Indeed, the fact of not detecting a fade event might harm the feeder link. On the other hand, we have to monitor the FP rate in order to keep the number of SGD switches reasonable as suggested in \cite{ventouras_assessment_2021}. 

\begin{figure}[h]
	\centering
	\begin{subfigure}{\linewidth}
		\centering
		\includegraphics[width=\sizing{0.3}{0.58}\linewidth]{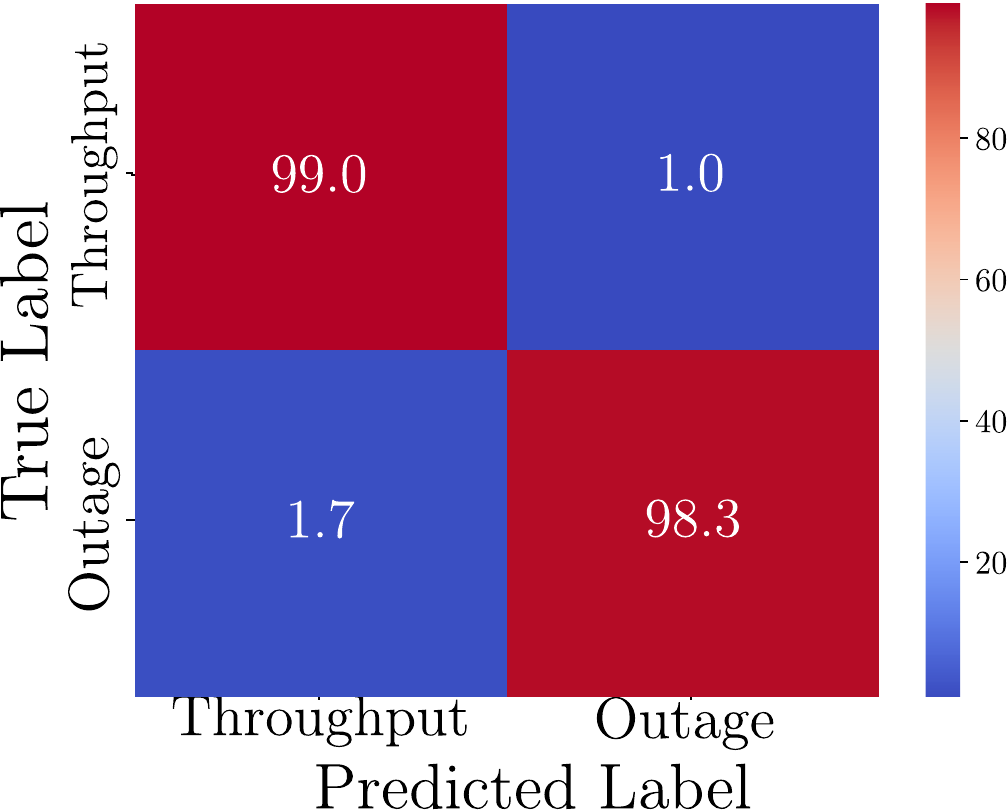}
		\caption{Confusion matrix with \airisp.}
		\label{fig:cm_lstm}
	\end{subfigure}
	\begin{subfigure}{\linewidth}
		\centering
		\includegraphics[width=\sizing{0.3}{0.58}\linewidth]{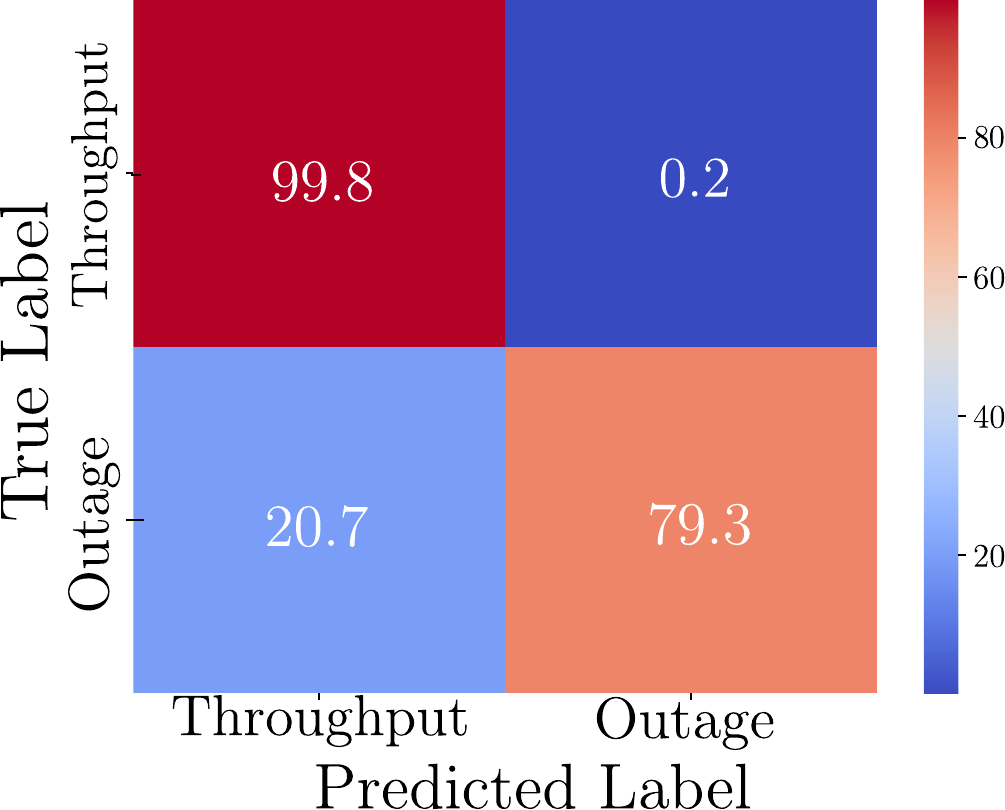}
		\caption{Confusion matrix with PH.}
		\label{fig:cm_persistence}
	\end{subfigure}
	\caption{Results $\Delta_t=90~\mathrm{s}$ and $\alpha=5~\mathrm{dB}$ with $T_1$ dataset.}
	\label{fig:cm}
	\vspace{-5mm}
\end{figure}

One example of result is shown in Fig. \ref{fig:cm} in the form of confusion matrices (in percents) obtained by running both algorithms on the $T_1$ database for $\Delta_t=90~\mathrm{s}$ and $\alpha=5~\mathrm{dB}$. 
 
 It appears that the number of FN \jcmm{significantly decreases} by using of our algorithm ($1.7\%)$ with respect to the PH ($20.7\%$). 
 For the case presented in Fig. \ref{fig:cm_persistence} a few FP ($0.2\%$) were found, which \jcmm{was} expected since the PH yields conservative triggering. In other terms, the lack of anticipation of \jcmm{PH introduces} more $\eventph$ with zero values that explains the \jcmm{low probability} of FP.
 Nevertheless, the FP rate for \airis remains moderate ($1\%$), this slight increase comes with no surprise because the LSTM architecture captures the dynamics of the system and thus can erroneously trigger the SGD.
 Overall, for this $(\alpha,\Delta_t)$ performed over $T_1$, the performances of \airis are clearly proving its interest because it is able to correctly anticipate strong fade events with respect to the simplistic PH. However, we need to validate this FN minimization performance trend at various locations and for several $(\alpha,\Delta_t)$ settings.
 \vspace{-5mm}
\subsection{Parametric sensitivity}
\label{ss:paramsens}
Let us assess the comparison between \airis and PH performances over parametric variation. In this subsection, we consider the $T_1$ dataset and run the algorithms for $\alpha \in \{5,10,15,20\}\mathrm{dB}$ and $\Delta_t \in \{30,60,90,120\}\mathrm{s}$. These spans of values are motivated by practical considerations : i) the SGD is not relevant for \jcmm{moderate} fade events (below $5\mathrm{dB}$) since other FMT are applicable in this case ii) the SGD can be reconfigured under few tens of seconds. Predictions exceeding two minutes are outside the scope of the \jcmm{present} paper since they would require extrinsic information (\textit{e.g.}, weather prediction data) to be performed.      

\begin{figure}[h]
	\centering
	\includegraphics[width=\sizing{0.5}{1.0}\linewidth]{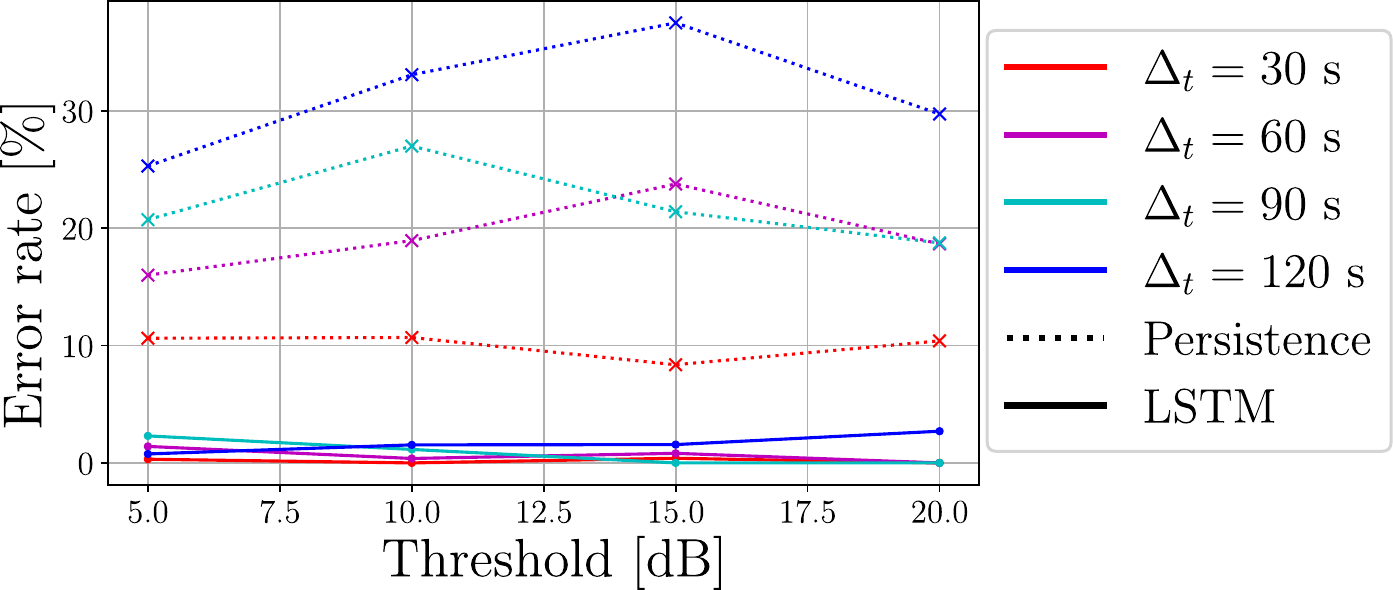}
	\caption{False negative rates with respect to $\Delta_t$ and $\alpha$, for PH and \airis predictors (\jcmm{$T_1$ experimental data}).}
	\label{fig:fn}
\end{figure}

The results for the FN are shown in Fig. \ref{fig:fn} for Q/V band data \jcmm{measured} in Toulouse ($T_1$). This result has been plotted using $16$ \airis training processes. The runs over the test database yielded FN rates under $2\%$ for each $(\alpha,\Delta_t)$ for our algorithm while the minimal FN rate for PH is close to $10\%$ and, \jcmm{as expected}, deteriorates as $\Delta_t$ increases. The performances in terms of FP for \airis (not shown) are under $2\%$, which remains reasonable. Therefore, we highlighted the robustness of our algorithm for a wide range of realistic SGD \jcmm{parameterizations}.  

\vspace{-4mm}
\subsection{Geographic and frequency sensitivity}

\begin{figure}[h]
	\centering
	\includegraphics[width=\sizing{0.5}{1.0}\linewidth]{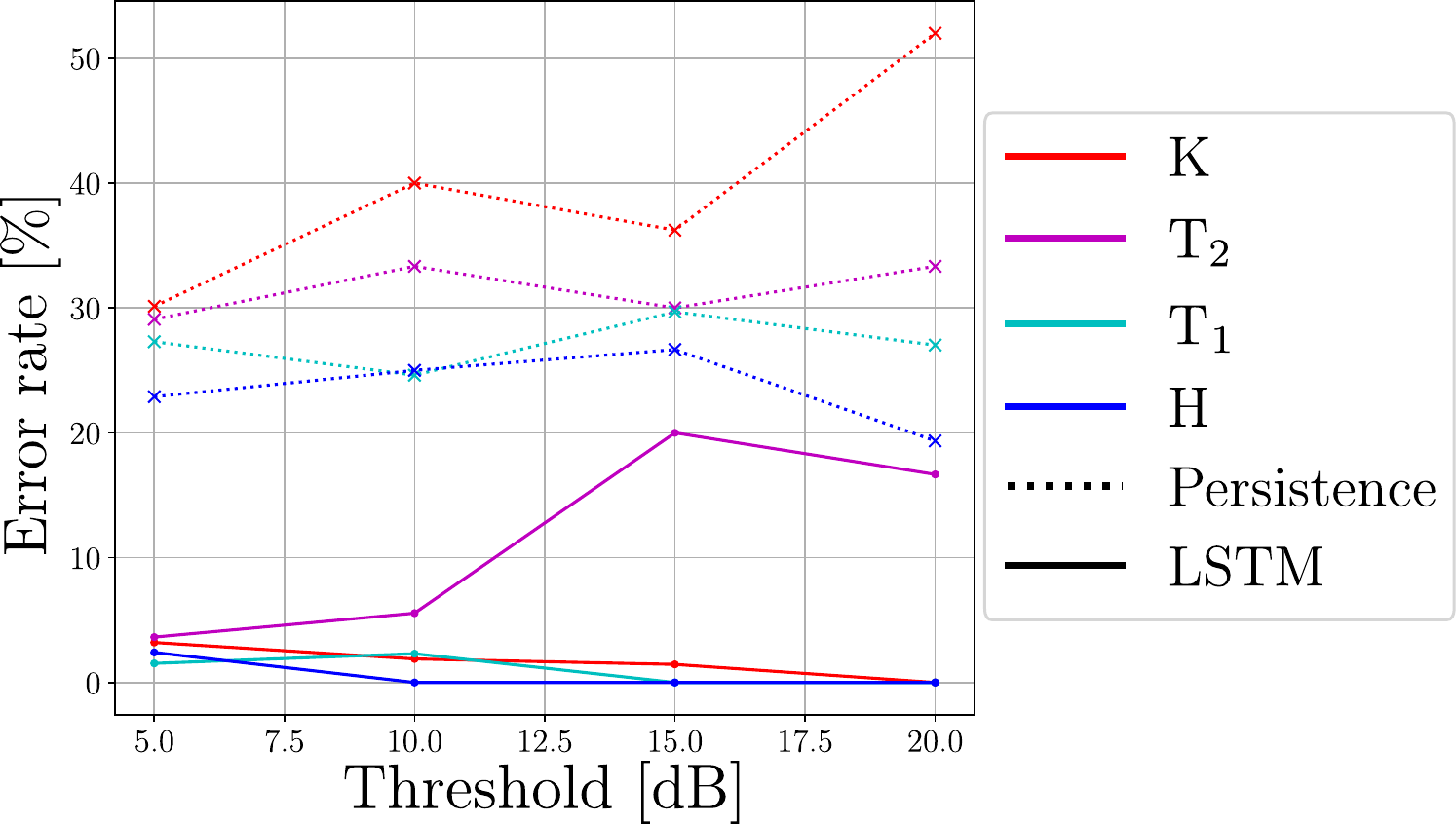}
	\caption{False negative rates with respect to $\alpha$ at different locations, for PH and \airis predictors $\Delta_t=120~\mathrm{s}$.}
	\label{fig:fn_geo}
\end{figure}

In this subsection, we perform the same tests than in Section \ref{ss:paramsens} over the \jcmm{four} datasets presented in Table \ref{tab:data}. \jcmm{For the sake of clarity}, we fixed $\Delta_t=120~\mathrm{s}$ since this horizon would be the more challenging for \airisp. The results for the FN are shown in Fig. \ref{fig:fn_geo}. 

First, we clearly see an improvement for the FN rates between \airis and PH performances for the $K$ dataset in red, which consists in more severe tropospheric events \jcmm{(tropical area)}. Indeed, \jcmm{convective rain cells, typical at such latitudes, introduce rapid increases in attenuation that remain} undetected by the simple threshold of the PH. This fact come without any surprise because the LSTM cells are able to capture the attenuation dynamics \jcmm{in the short-term}. This behavior of \airis seems to be generalized to the other datasets, except for $T_2$. For this latter, the performances for Toulouse at Ka band (in purple) seems to be deteriorated when $\alpha>15\mathrm{dB}$. This is explainable by the lack of such severe rain events at a lower frequency than in $T_1$ ($20\mathrm{GHz}$ \textit{vs.} $40\mathrm{GHz}$) in a temperate climate region. Overall, this fact reminds us the importance of having significant datasets to guarantee optimal performances. 

\section{Conclusion}

In this paper, we proposed a novel SGD switching algorithm based on LSTM deep learning architecture called \airis. Despite having a relatively small number of parameters, \airis has been validated by comparing it to a baseline algorithm for a wide span of SGD parameters and at different locations and frequencies, making it an easily deployable solution to enhance VHTS communications. Future work will include low elevations and extrinsic measurements in order to make \airis more adaptive and enhance its performances.    

\section*{Acknowledgments}
The authors wish to thanks Pr. Bousquet and Dr. Castanet for their precious pieces of advice and Dr. Monvoisin for providing the total attenuation dataset. 

\balance
\bibliographystyle{IEEEtran}
\bibliography{biblio}

\end{document}